\documentclass[aps,prd,amsmath,amsfonts,amssymb,eqsecnum,nofootinbib,
  floatfix,secnumarabic,preprintnumbers,superscriptaddress,nobalancelastpage,
  twocolumn]{revtex4}
\usepackage[utf8]{inputenc}
\usepackage{color,graphicx,multirow,slashed,hyperref}
\usepackage[export]{adjustbox}
\hypersetup{bookmarks=true,unicode=true,pdftoolbar=true,pdfmenubar=true,
  pdffitwindow=false,pdfstartview={FitB},pdftitle={HHH-ditau},
  pdfauthor={B.~Fuks, J.H.~Kim, S.J.~Lee}, pdfsubject={Subject},
  pdfcreator={Creator},pdfproducer={Producer}, pdfkeywords={Higgs properties}{
  Beyond the Standard Model Physics}{100 TeV proton-proton collider},
  pdfnewwindow=true,colorlinks=true,
  linkcolor=blue,citecolor=magenta,filecolor=magenta,urlcolor=cyan}

\newcommand {\beq} {\begin{equation*}}
\newcommand {\eeq} {\end{equation*}}

\definecolor{red1}{cmyk}{0,1,1,0.1}

\newcommand{\met}{\slashed{E}_T\,}

\newcommand{\lhhh}{\lambda_{hhh}^{\rm SM}}
\newcommand{\lhhhh}{\lambda_{hhhh}^{\rm SM}}

\newcommand{\amc}{{\sc MadGraph5\textunderscore}a{\sc MC@NLO}}
\newcommand{\frules}{{\sc Feyn\-Rules}}

\newcommand{\cf}{{\it cf. }}

\begin{document}

\date{\today}

\title{Scrutinizing the Higgs quartic coupling\\
   at a future 100 TeV proton-proton collider with taus and b-jets}

\author{Benjamin Fuks}
\affiliation{Sorbonne Universit\'es, Universit\'e Pierre et Marie Curie
  (Paris 06), UMR 7589, LPTHE, F-75005 Paris, France}
\affiliation{CNRS, UMR 7589, LPTHE, F-75005 Paris, France}
\affiliation{Institut Universitaire de France, 103 boulevard Saint-Michel,
  75005 Paris, France}

\author{Jeong Han Kim}
\affiliation{Department of Physics and Astronomy, University of Kansas, Lawrence, KS, 66045, USA}

\author{Seung J.~Lee}
\affiliation{Department of Physics, Korea University, Seoul 136-713, Korea}
\affiliation{School of Physics, Korea Institute for Advanced Study, Seoul 130-722, Korea}

\begin{abstract}

The Higgs potential consists of an unexplored territory in which the electroweak
symmetry breaking is triggered, and it is moreover directly related to the
nature of the electroweak phase transition. Measuring the Higgs boson cubic
and quartic couplings, or
getting equivalently information on the exact shape of the Higgs potential, is
therefore an essential task. However, direct measurements beyond the cubic
self-interaction of the Higgs boson consist of a huge challenge, even for a
future
proton-proton collider expected to operate at a center-of-mass energy
of 100~TeV. We present a novel approach to extract model-independent constraints
on the triple and quartic Higgs self-coupling by investigating triple
Higgs-boson hadroproduction at a center-of-mass energy of 100~TeV, focusing on
the $\tau \tau b \bar{b} b \bar{b}$ channel that was previously overlooked due
to a supposedly too large background. It is thrown into sharp relief that the
assist from transverse variables such as $m_{T2}$ and a boosted configuration
ensures a high signal sensitivity. We derive the luminosities that would be
required to constrain given deviations from the Standard Model in the
Higgs self-interactions, showing for instance that a $2\sigma$ sensitivity
could be achieved for an integrated luminosity of 30~ab$^{-1}$ when Standard
Model properties are assumed. With the prospects of combining these findings
with other triple-Higgs search channels, the Standard Model Higgs quartic
coupling could in principle be reached with a significance beyond the $3\sigma$
level.
\end{abstract}
\maketitle

\section{Introduction}\label{sec:intro}
The discovery of a Higgs boson at the
Large Hadron Collider (LHC) accomplished the long waited physics goals of
getting hints on the nature of the electroweak symmetry breaking (EWSB)
mechanism and understanding the generation of the fermion
masses. While
the discovered Higgs boson appears to be highly compatible with the Standard
Model (SM) expectation~\cite{Khachatryan:2016vau}, current data is still
insufficient for revealing the true nature of the EWSB dynamics. Further pieces
of information related to the shape of the Higgs potential are indeed needed,
such as measurements of the Higgs cubic, quartic and even higher-order
self-couplings. This would furthermore allow us to investigate whether the
electroweak phase transition is of the first or second order, a fact related to
the matter-antimatter asymmetry in the universe as a strong first order
electroweak phase transition can potentially realize one of the Sakharov
conditions for baryogenesis. Measuring the Higgs cubic and quartic
self-couplings is consequently one of the major physics goals of the future
high-energy physics program.

Di-Higgs production via gluon fusion offers the first playground to access the
Higgs cubic coupling, in particular within the high-luminosity phase
of the LHC expected to collect an integrated luminosity of 3~ab$^{-1}$ of
data at a center-of-mass energy of 14~TeV~\cite{Dolan:2012rv,Baglio:2012np}. The
associated sizable (SM) production cross section of about
43~fb~\cite{deFlorian:2015moa} allows one to make use of various final states to
probe the Higgs cubic coupling, the two most promising signatures
relying on final state systems made of four $b$-jets, or of a pair of photons
and either a pair of $b$-jets or tau leptons~\cite{ATL-PHYS-PUB-2016-024,%
ATL-PHYS-PUB-2017-001,CMS:2015nat}. At a future proton-proton collider aiming to
operate at a center-of-mass energy of 100~TeV, the $b \bar{b} \gamma \gamma$
channel keeps its leading role and measurements at a precision of about $3-4\%$
could be expected for a luminosity of 30~ab$^{-1}$~\cite{Contino:2016spe,%
Azatov:2015oxa}. None of these searches are, however, designed to probe the
Higgs quartic coupling.

\begin{figure}
  \centering
  \includegraphics[valign=t, width=.21\textwidth]{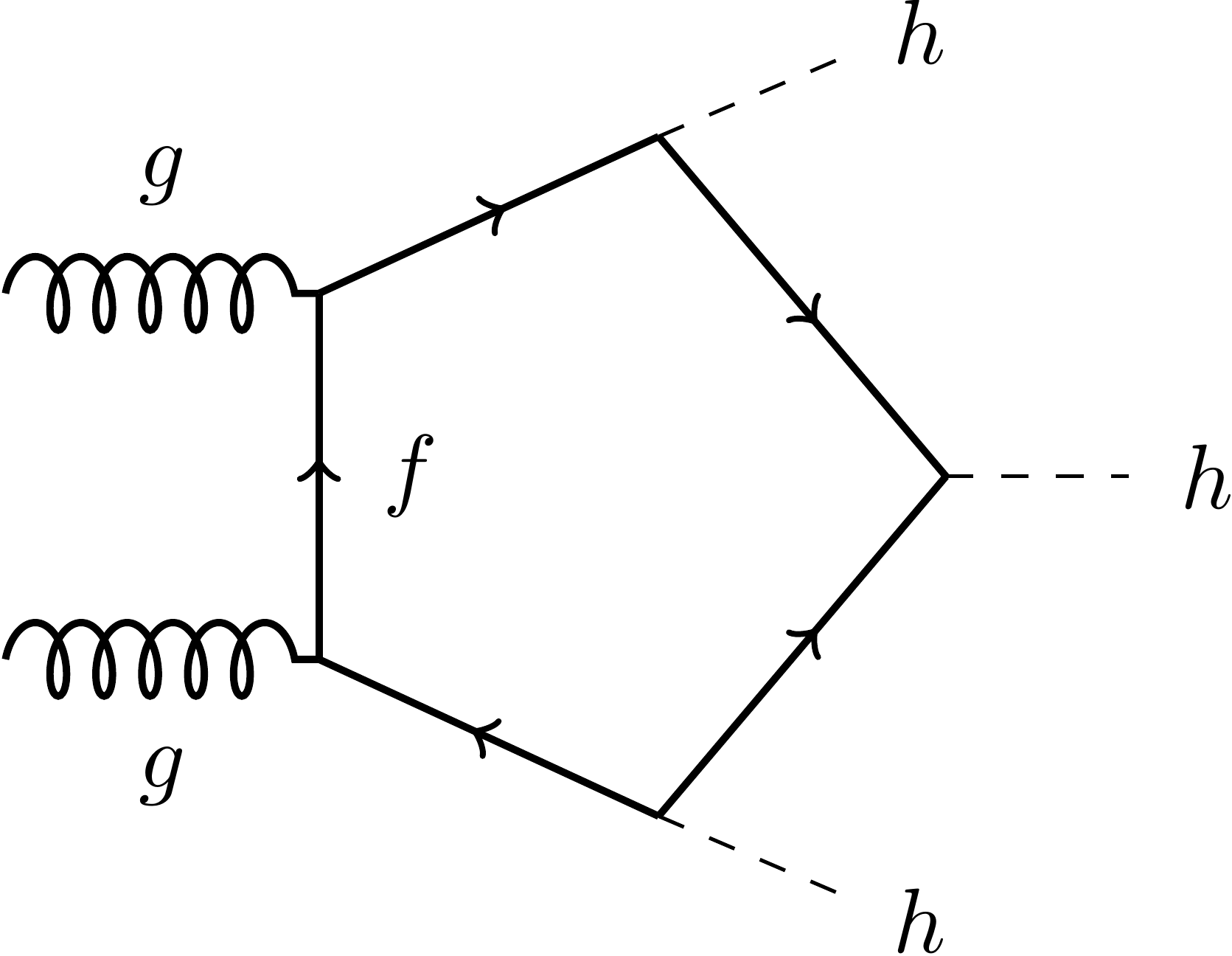}
  \includegraphics[valign=t, width=.21\textwidth]{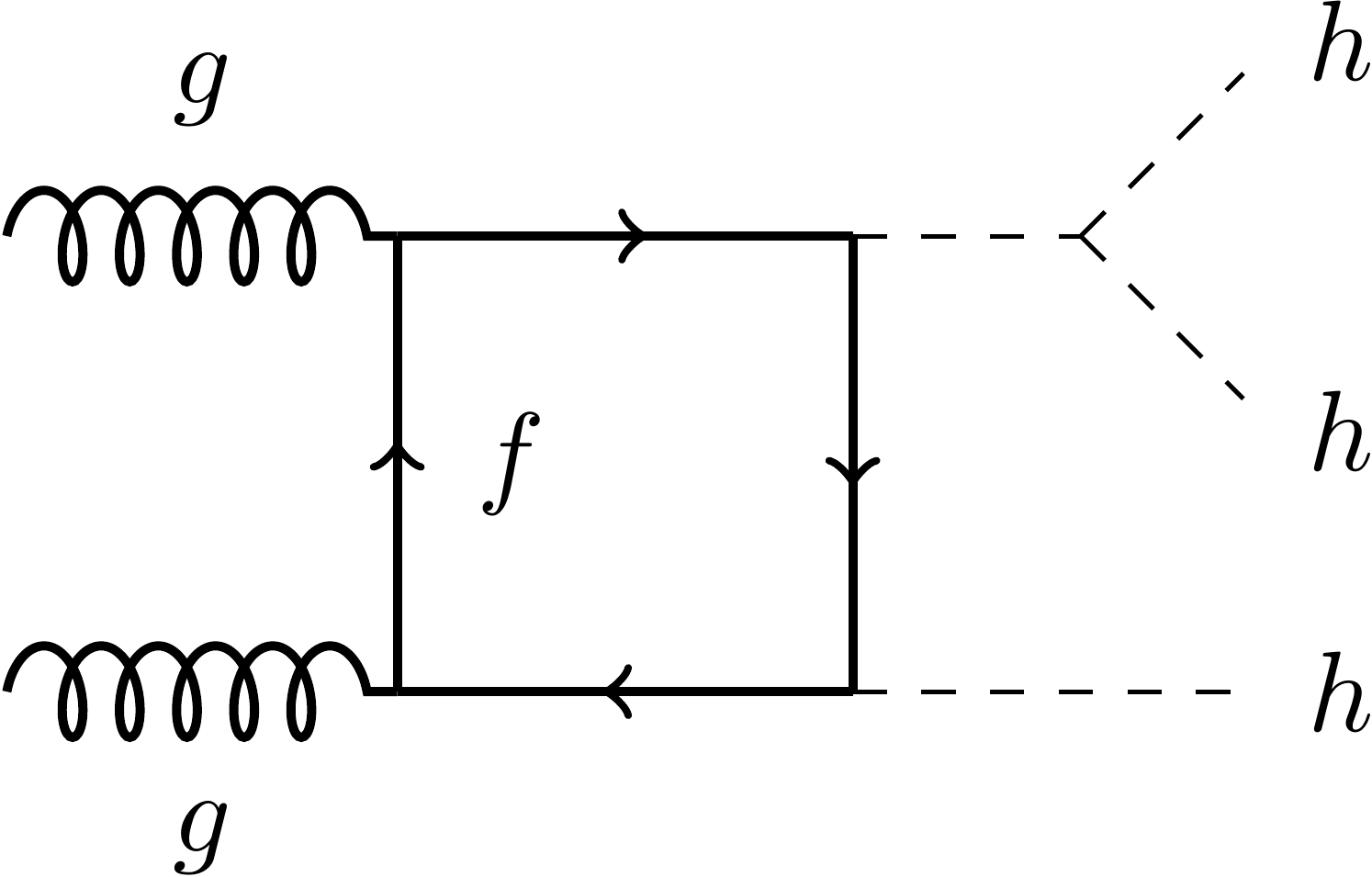}\\
  \includegraphics[valign=t, width=.21\textwidth]{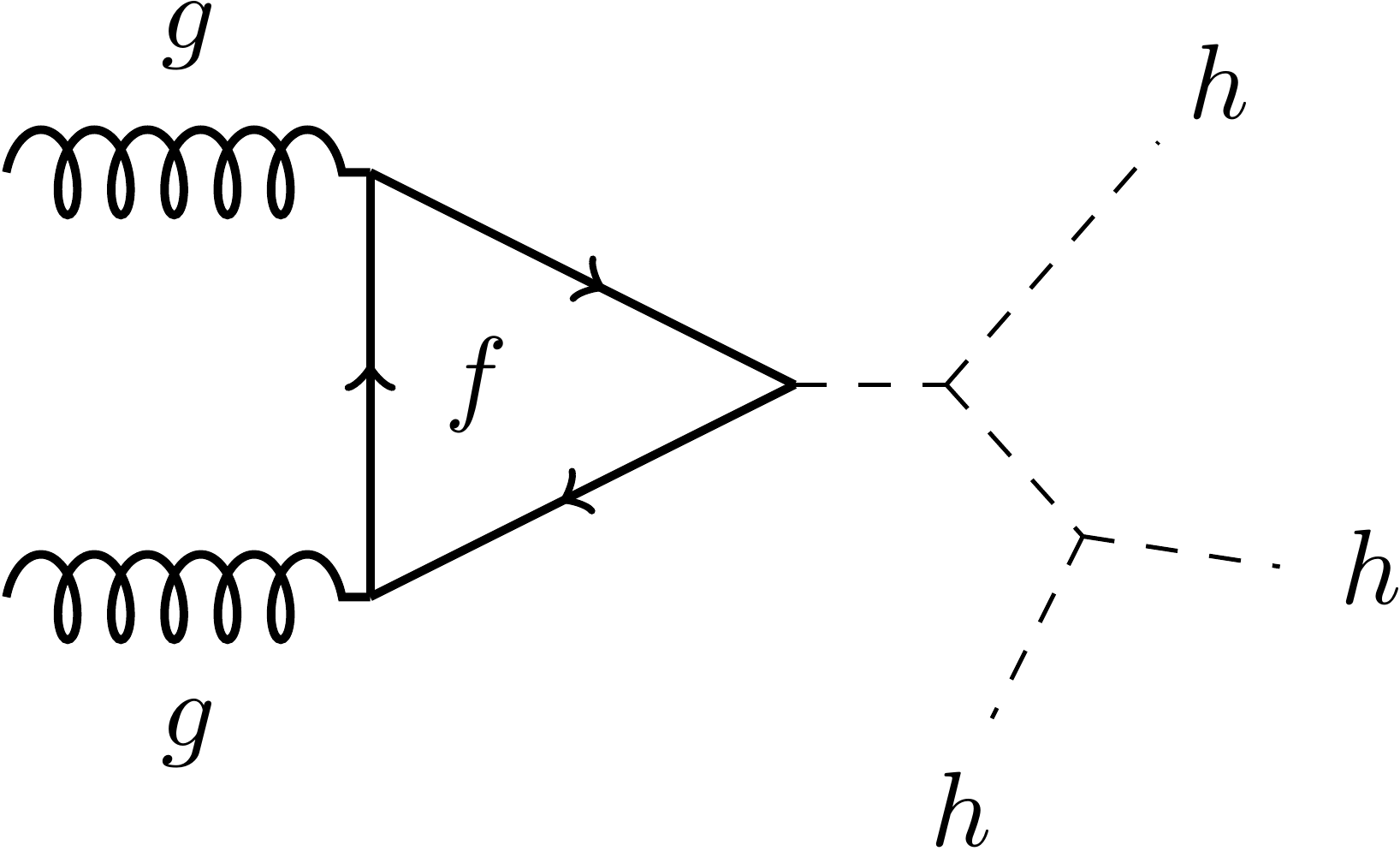}
  \includegraphics[valign=t, width=.21\textwidth]{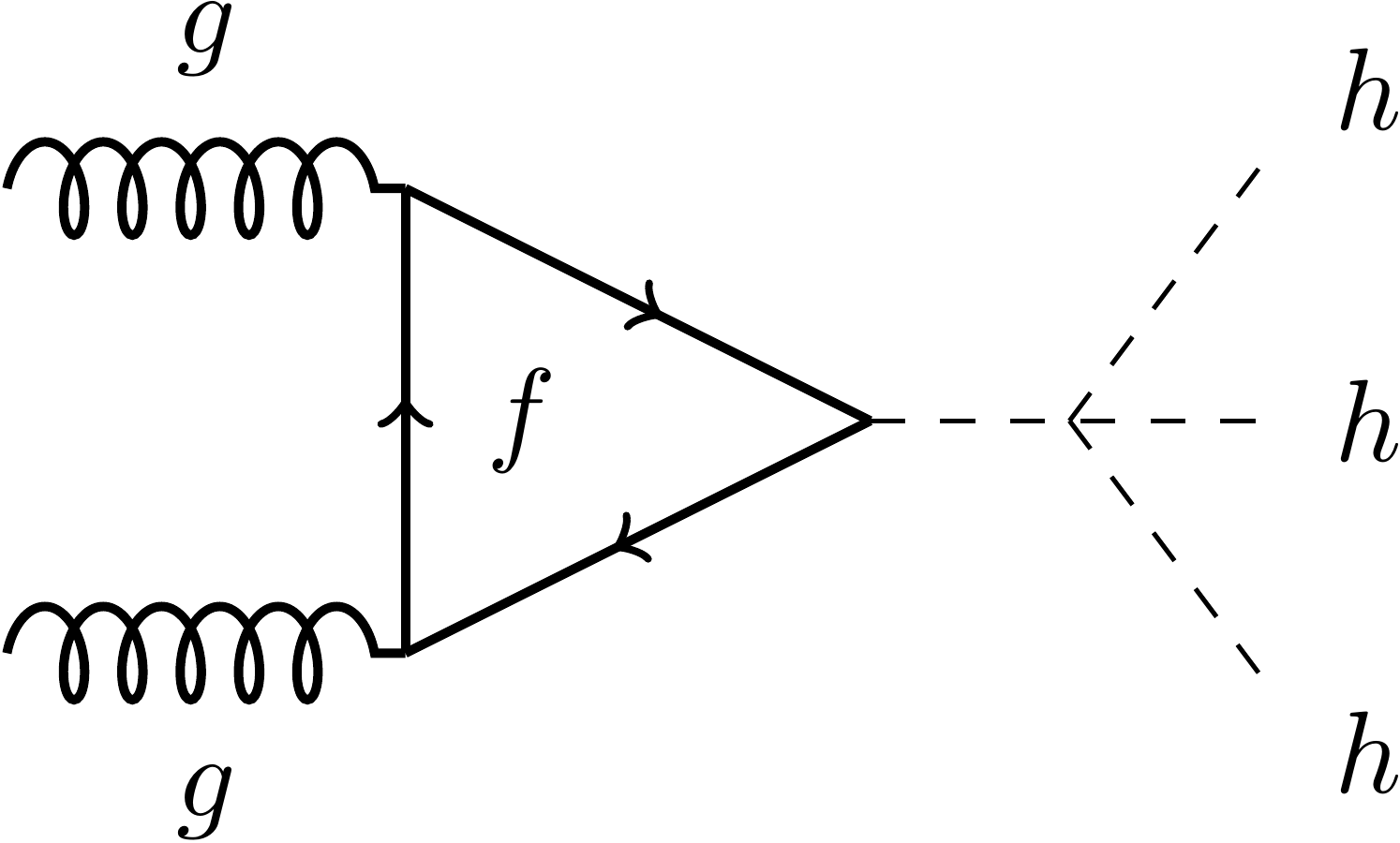}
  \caption{\it Representative leading-order Feynman diagrams for triple
   Higgs production in proton-proton collisions.}
  \label{fig:Diagram}
\end{figure}

In the SM, triple-Higgs production mostly arises, at the leading-order in QCD,
by gluon fusion (see Fig.~\ref{fig:Diagram}). Such a process faces a rather grim
prospect at the LHC, mainly because of a small signal rate of
${\cal O}(0.1)$~fb~\cite{Plehn:2005nk,Binoth:2006ym,Maltoni:2014eza}, so that
the study of this process is left to the experimental program of the
post-LHC era that is currently under discussion at CERN and
IHEP~\cite{Arkani-Hamed:2015vfh}. Feasibility analyses have so far shown that
the $b\bar b b\bar b \gamma\gamma$ channel can be used to constrain the size of
the quartic Higgs coupling in a model-independent way, the interaction strength
being allowed to deviate by a factor of at most ${\cal O}(10)$ from the SM after
considering an integrated luminosity of 30~ab$^{-1}$~\cite{Fuks:2015hna,%
Chen:2015gva,Papaefstathiou:2015paa}. The prospects of the $b\bar b W W W W$
decay mode have also been explored, and it was shown that a new physics
triple-Higgs signal is in principle detectable~\cite{Kilian:2017nio}.

In this article, we embark on reinvestigating triple Higgs production at a
100~TeV proton-proton collider to be more confident on the sensitivity of such a
machine to the quartic Higgs self-coupling. We focus on the
more challenging, branching-ratio-enhanced, $b\bar bb\bar b\tau^+\tau^-$
signature. Contrary to the $b\bar b b\bar b \gamma\gamma$ channel, it receives a
severe background contamination that yields a weaker expected
sensitivity~\cite{Fuks:2015hna}. However, with the effort of exploiting
previously overlooked advantages of the ditau system and a boosted
configuration, we show in this work that the $b\bar b b\bar b \tau\tau$ channel
can
be promoted to a leading discovery channel for triple-Higgs production.

This paper is organized as follows. In Sec.~\ref{sec:model}, we introduce the
adopted simplified model parameterizing in a model-independent way any new
physics effect on the Higgs self-interactions, and we present technical details
related to our simulation setup.
Sec.~\ref{sec:analysis} is dedicated to our event selection strategy and
exhibits details on its specificity. Our results are given in
Sec.~\ref{sec:results}, together with prospects for a future 100~TeV
proton-proton colliders.

\section{Theoretical framework and technical details}\label{sec:model}
In order to probe for possible new physics effects in multiple-Higgs
interactions, we modify in a model-independent fashion the SM Higgs potential,
\beq
  V_{\rm h} =  \frac{m_h^2}{2} h^2 + (1 + \kappa_3) \lhhh v h^3 +
    \frac14 (1 + \kappa_4) \lhhhh h^4 \ ,
\label{eq:vh}\eeq
by introducing two $\kappa_i$ parameters that vanish in the SM. In our
notation, $h$ denotes the physical Higgs-boson field, $m_h$ its mass and $v$ its
vacuum expectation value. The SM self-interaction strengths moreover read
\beq
  \lhhh = \lhhhh = \frac{m_h^2}{2 v^2} \ .
\eeq
We simulate our triple Higgs signal and the associated backgrounds by
implementing the above Lagrangian in the \frules~package~\cite{Alloul:2013bka}
that we use along with the {\sc NloCT} program~\cite{Degrande:2014vpa} to
generate a UFO library~\cite{Degrande:2011ua}. The latter allows for event
generation for both tree-level and loop-induced processes within the
\amc~\cite{Alwall:2014hca,Hirschi:2015iia} framework, that we use to convolute
hard scattering matrix elements with the next-to-leading (NLO) set of NNPDF 2.3
parton densities~\cite{Ball:2012cx} for a center-of-mass energy of
$\sqrt{s}=100$~TeV. The hard-scattering events are then decayed, showered and
hadronized within the {\sc Pythia 6} environment~\cite{Sjostrand:2006za} and
reconstructed by using the anti-$k_T$ algorithm~\cite{Cacciari:2008gp} as
implemented in {\sc FastJet}~\cite{Cacciari:2011ma}, with a radius of
$R = 1$ and 0.4 for a fat jet and slim jet definition, respectively.

Hadronic taus are defined as specific slim jets for which there is no hadronic
object of \mbox{$p_T > 1$~GeV} and no photon with a \mbox{$p_T > 1.5$~GeV} at an
angular distance of the jet axis greater than \mbox{$r_{\rm in}=0.1$} and
smaller than \mbox{$r_{\rm out}=0.4$}. The resulting tau-tagging efficiency is
of about 50\%, for a fake rate of mistagging a light-flavor jet as a tau of
roughly 5\%. Those performances can be compared to what could be expected from
the high-luminosity phase of the LHC, for which an efficiency of 55\% can be
expected for a mistagging rate of 0.5\%~\cite{CMS:2015nat}.

Our analysis relies on the reconstruction of boosted Higgs bosons. To
this aim, we employ the template overlap method~\cite{Almeida:2010pa,%
Almeida:2011aa} as embedded in the {\sc TemplateTagger}
program~\cite{Backovic:2012jk}, and we use a new template observable
derived from the $ty$ quantity proposed in Ref.~\cite{Kim:2016plm}, which we
here maximize over the different three-body Higgs templates. We make use
of various two-body and three-body
(NLO) Higgs templates featuring a sub-cone size of 0.1 to compute the
discriminating overlaps $Ov_2^h$ and $Ov_3^h$, respectively, that allow for a
boosted Higgs boson identification. The performance of the method yields
a tagging efficiency of 40\% for a mistagging rate of 2\%.

\begin{figure}
  \centering
  \includegraphics[scale=0.55]{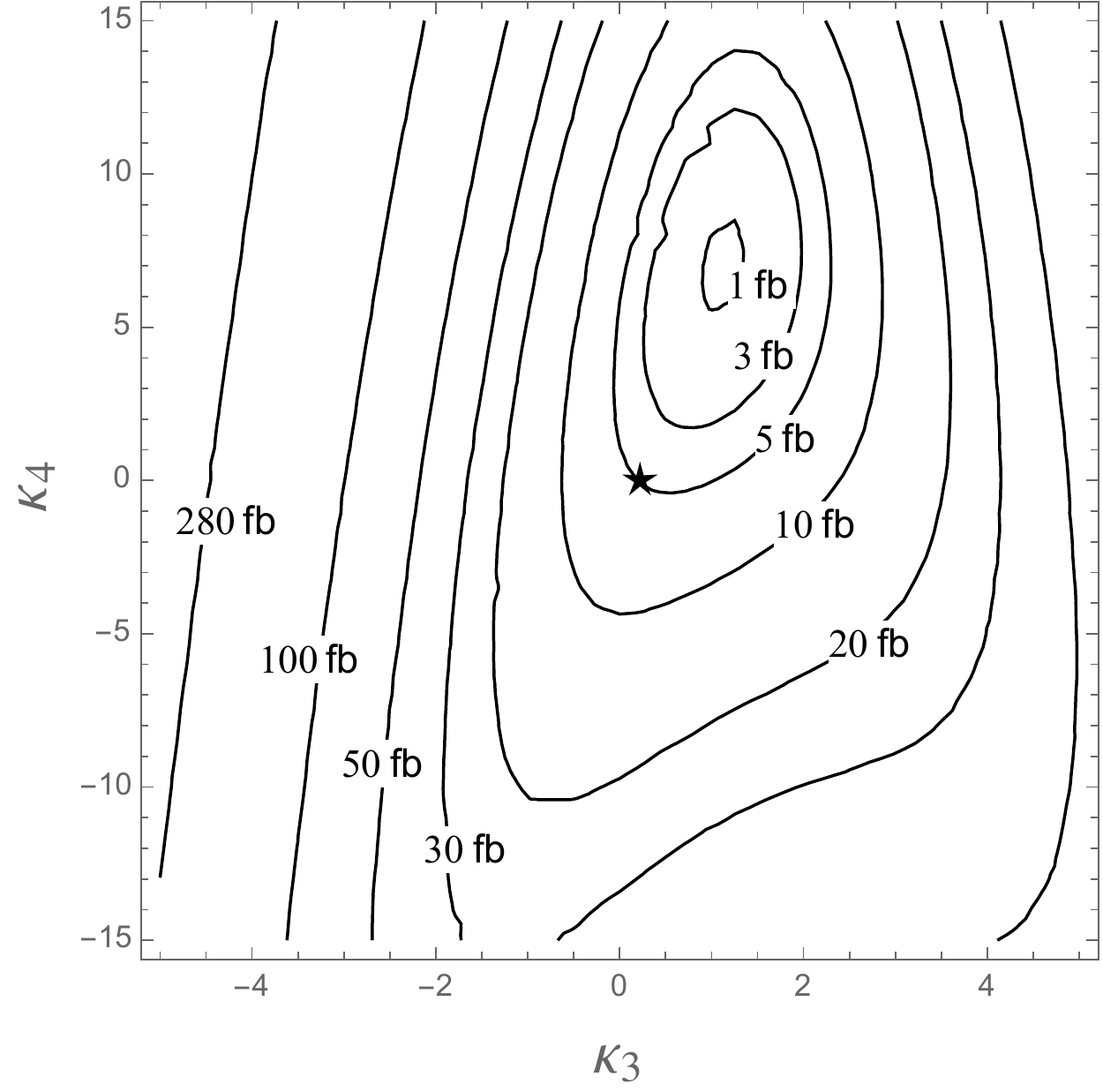} 
  \caption{\it Triple-Higgs production cross-section for a center-of-mass energy
   of $\sqrt{s} = 100$~TeV presented as a function of the $\kappa_3$ and
   $\kappa_4$ parameters depicting the possible deviations from the SM
   (indicated by a black star). The results include a conservative NLO
   $K$-factor of 2.}
  \label{fig:Xsection}
\end{figure}

As suggested by the representative Feynman diagrams of Fig.~\ref{fig:Diagram},
triple-Higgs production depends on both $\kappa_i$ parameters
as well as on the top Yukawa coupling. While in either an effective field theory
framework or an ultraviolet-complete model building approach, the $\kappa_i$
parameters are not independent, they will be varied independently in our study.
Moreover, the top Yukawa coupling is assumed to be fixed to its SM value. The
resulting production cross section is presented in Fig.~\ref{fig:Xsection} in
the $(\kappa_3, \kappa_4)$ plane after including a flat NLO $K$-factor of
2~\cite{deFlorian:2016sit}. The
sign of the $\kappa_3$ parameter turns out to be crucial due to respective
constructive and destructive interference patterns when $\kappa_3$ is negative
and positive. As a consequence, the cross section can be reduced to below the
fb level when both $\kappa$ parameters are positive (and not too large), making
this corner of the parameter space hard to probe. The variations in $\kappa_4$
are in addition mild for any fixed value of $\kappa_3$, so that only poor
constraints could be expected from any potential measurement.

Among all triple-Higgs production signatures, we make use of the $b\bar bb\bar
b\tau^+\tau^-$ channel with two hadronic tau decays to probe deviations in
the Higgs self-interactions. Whilst the branching ratio is large ($\sim 6.3\%$),
the background contamination is expected to be important~\cite{Fuks:2015hna}. We
however demonstrate in the next sections that
previously overlooked advantages stemming from the usage of specific kinematic
properties of the ditau systems and the potentially boosted configuration of the
$b$-jet pairs could largely increase the signal significance.

\begin{table}
  \renewcommand{\arraystretch}{1.1}
  \setlength\tabcolsep{6pt}
  \begin{tabular}{c|c|c}
   Class & Backgrounds & Cross section [ab] \\
   \hline
   \hline
   \multirow{5}{*}{$t/W \; \rm samples$}
      & $t_{\tau} \bar{t}_{\tau} h_{b \bar{b}}$ & $2.3 \times 10^4$\\
      & $t_{\tau} \bar{t}_{\tau} Z_{b \bar{b}}$ & $6.6 \times 10^3$\\
      & $t_{\tau} \bar{t}_{\tau} b \bar{b}$     & $4.7 \times 10^5$\\
      \cline{2-3}
      & $W^{+}_{\tau}W^{-}_{\tau} b\bar{b}b\bar{b}$ & $4.7 \times 10^5$\\
      \cline{2-3}
      & $t \bar{t} t \bar{t}$ & $6.6 \times 10^4$\\
   \hline
   \multirow{9}{*}{$X_{\tau \tau} \; + \; \rm{jets} $}
      & $X_{\tau \tau} b \bar{b} b \bar{b}$ & $6.9 \times 10^4$\\
      & $X_{\tau \tau} b \bar{b} j j $      & $1.5 \times 10^7$\\
      \cline{2-3}
      & $X_{\tau \tau} t_{h} \bar{t}_{h} $  & $1.6 \times 10^5$\\
      \cline{2-3}
      & $X_{\tau \tau} Z_{b \bar{b}} b \bar{b}$ & $2.0 \times 10^3$\\
      & $Z_{\tau \tau} h_{b \bar{b}} b \bar{b}$ & $300$\\
      \cline{2-3}
      & $X_{\tau \tau} Z_{b \bar{b}} Z_{b \bar{b}}$  & $23$\\
      & $Z_{\tau \tau} h_{b \bar{b}} Z_{b \bar{b}}$  & $15$\\
      \cline{2-3}
      & $h_{\tau \tau} h_{b \bar{b}} Z_{b \bar{b}}$  & \multirow{2}{*}{$11$}\\
      & $h_{b \bar{b}} h_{b \bar{b}} Z_{\tau \tau}$  &\\
      \hline
    \multirow{1}{*}{Di-Higgs}
      & $h_{\tau \tau} h_{b \bar{b}} + \rm jet $ & $1.3 \times 10^3$\\
  \end{tabular}
  \caption{\it Fiducial cross sections of all components of the SM background
    after the baseline selection described in Sec.~\ref{sec:analysis}. The
    results include an NLO $K$-factor of 2, and the
    suffixes `$\tau$' and `$b \bar{b}$' respectively indicate decays into a
    tau-lepton and a $b \bar{b}$ pair, $t_h$ denoting similarly a
    hadronically-decaying top quark.}
  \label{tab:BGD}
\end{table}

The various components of the SM background can be classified into three
categories regarding their response to the basic selection criteria introduced
in Sec.~\ref{sec:analysis}. We denote by $t/W \; \rm samples$ the ensemble
of background processes featuring a top quark or a $W$-boson pair that decays
into a tau-enriched final state, together with the four-top background
contributions. The second class of SM backgrounds consists of the
$X_{\tau \tau} + \rm{jets}$ category with $X$ being a virtual photon, Higgs or
$Z$-boson decaying into a pair of tau leptons. Di-Higgs production in
association with jets finally forms the last class of background
processes on its own. The full list of considered SM backgrounds is summarized
in Table~\ref{tab:BGD}, where we additionally present the fiducial cross
sections, multiplied by a conservative NLO K-factor of 2, obtained after
requiring the presence of two hadronic taus and missing
transverse energy (\cf the baseline selection described in
Sec.~\ref{sec:analysis}).

\section{Signal selection}\label{sec:analysis}
Our triple-Higgs analysis relies for its baseline selection on the properties of
the $b\bar bb\bar b\tau^+\tau^-$ final state. We preselect events featuring
exactly two hadronic taus with a \mbox{$p_T>25$~GeV} and a
pseudorapidity \mbox{$|\eta|< 2.5$}, as well as a missing transverse energy
\mbox{$\met > 25$~GeV}.

\begin{figure}
  \centering
  \includegraphics[scale=0.4]{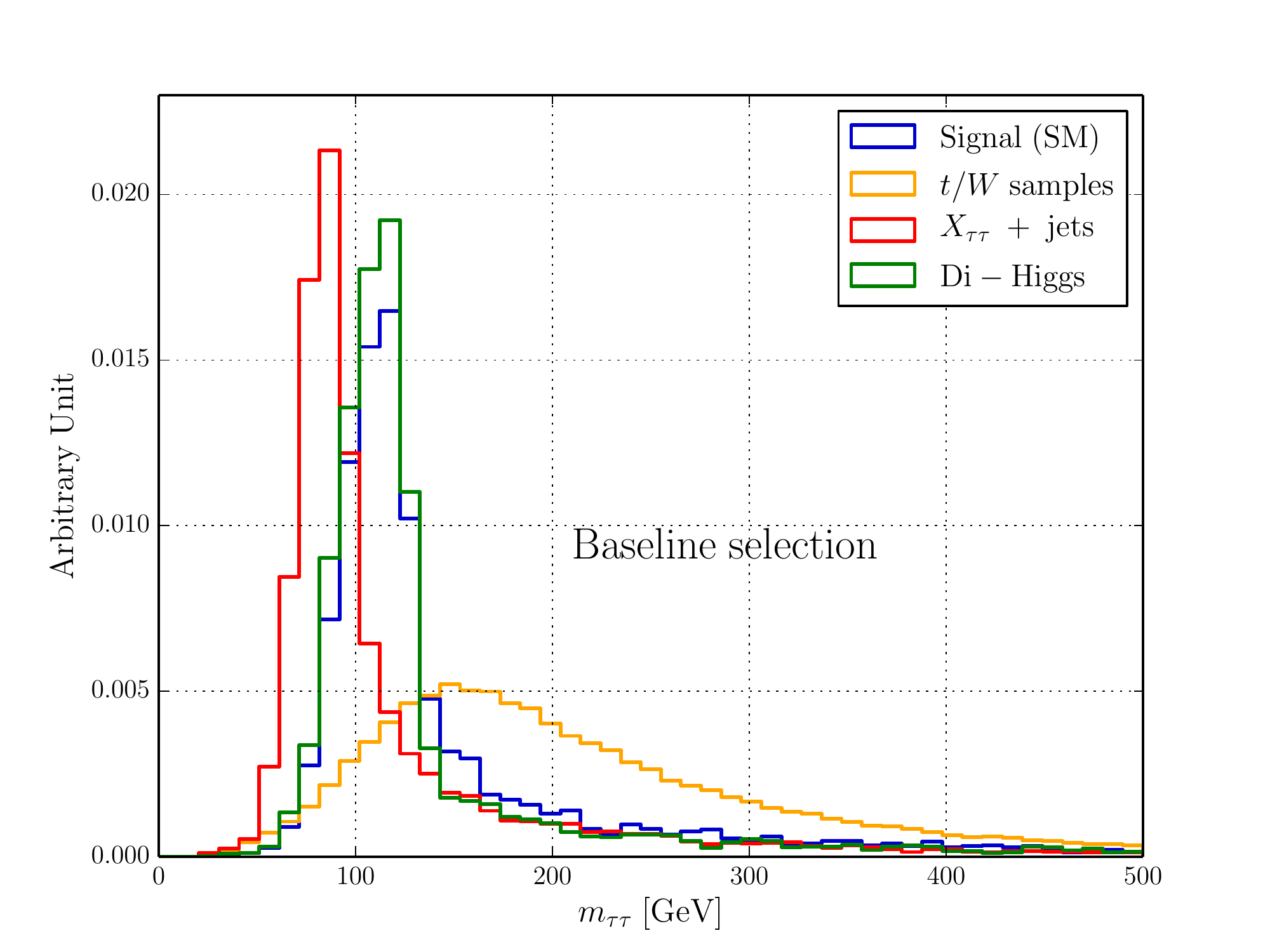}
  \caption{\it Distribution in the $m_{\tau\tau}$ invariant mass (as defined in
    the text) after the baseline selection for the three SM background
    categories and for a SM triple-Higgs signal.}
  \label{fig:mditau}
\end{figure}

After this preselection, the two taus are enforced to be compatible with the
decay of a Higgs boson by means of the $m_{\tau\tau}^{\rm Higgs-bound}$
and $m_T^{\rm True}$ variables~\cite{Barr:2011he,Barr:2009mx,Barr:2013tda}. The
former quantity is defined by minimizing, over all possible assignments for the
neutrino four-momenta, the invariant mass of the system made of the two tau jets
and the two invisible neutrinos. This minimization procedure however requires
that each tau jet is matched with a neutrino and that the resulting two-body
invariant mass is compatible with the tau mass. For cases for which there is no
such a solution, the $m_T^{\rm True}$ variable is constructed instead in the
same way, but without this last constraint. We present the resulting
$m_{\tau\tau}$ distribution in Fig.~\ref{fig:mditau},
$m_{\tau\tau}$ generically denoting $m_{\tau\tau}^{\rm Higgs-bound}$ when it can
be constructed and $m_T^{\rm True}$ otherwise. Most signal events exhibit an
$m_{\tau\tau}$ value lying between the $Z$ and the Higgs boson masses, whereas
background events from the $X_{\tau \tau} \; + \; \rm{jets}$ category mainly
feature smaller $m_{\tau\tau}$ values. We therefore impose that
\mbox{$m_{\tau\tau} \in [105, 135]$~GeV} to ensure compatibility with a Higgs
ditau
decay and a very good discrimination from the $X_{\tau \tau} \; + \; \rm{jets}$
background category.

We move on with the reconstruction of the two other Higgs bosons for which we
rely on a configuration where one of them is boosted and the other one is
resolved. We select events featuring at least one fat jet whose basic properties
satisfy \mbox{$p_T > 300$}~GeV and \mbox{$|\eta |< 2.5$}. The fat jet invariant
mass is moreover required to lie in the $[105, 135]$~GeV window and the template
overlaps are constrained to \mbox{$Ov_3^h > 0.7$} and \mbox{$Ov_2^h > 0.2$}. We
additionally require the presence of at least two slim jets and tag two of them
as candidates for a non-boosted Higgs decay. This tagging is such that the dijet
invariant mass \mbox{$m_{\rm jj}\in [105,135]$~GeV} minimizes
\mbox{$|m_{\rm{j j}} - m_h|$}.
Furthermore, one of the two tagged slim jets must be $b$-tagged and the fat
jet must contain a doubly-$b$-tagged substructure when we assume a $b$-tagging
efficiency of 70\% when a $B$-hadron is present in a cone of radius
\mbox{$R=0.4$} around the jet direction, for a corresponding mistagging rate of
1\%.

\begin{figure}
  \centering
  \includegraphics[scale=0.4]{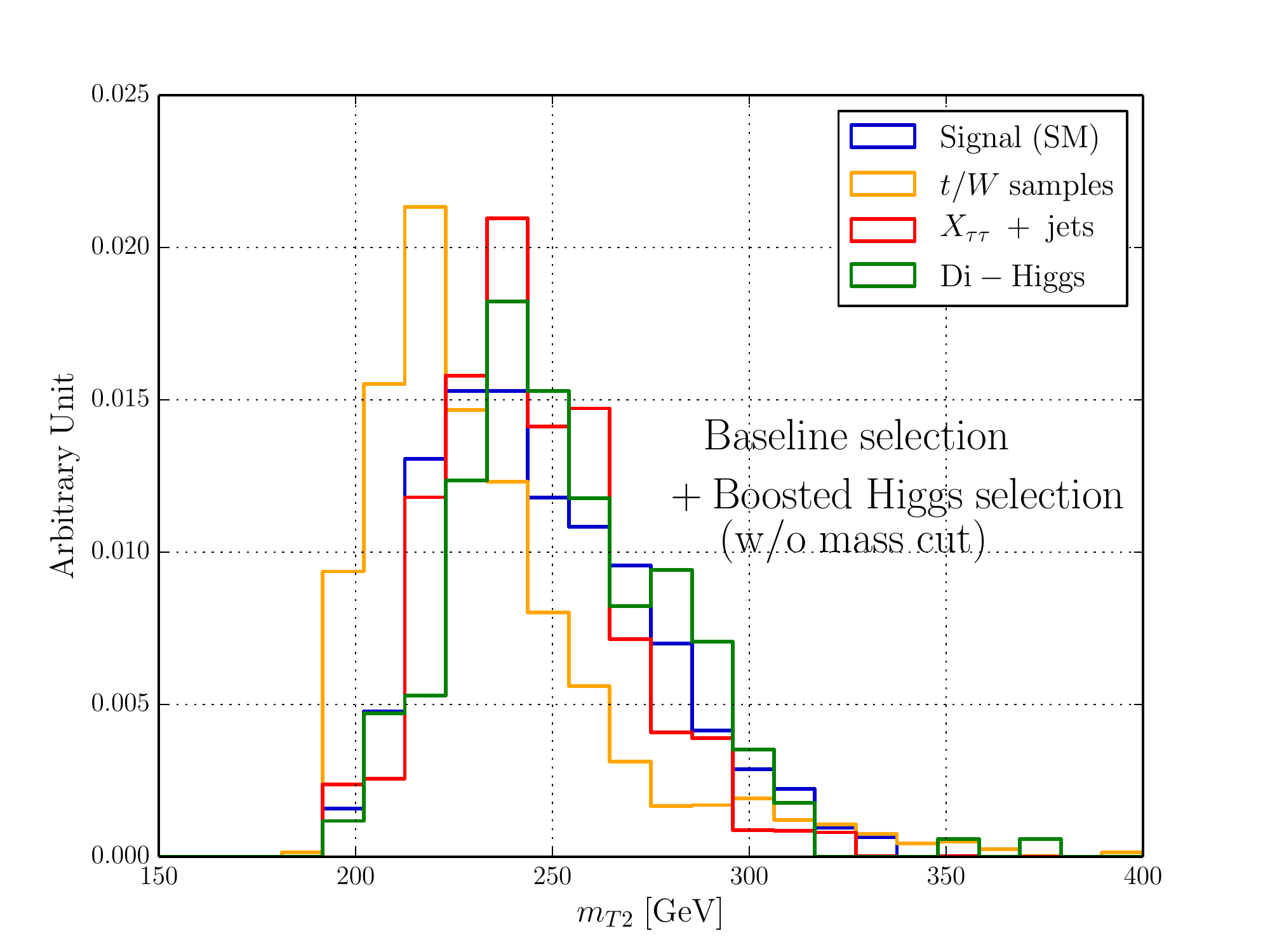}
  \caption{\it $m_{T2}$ spectra for the signal (in the case of a SM Higgs
    potential) and the various components of the background.}
 \label{fig:mt2}
\end{figure}

At this stage, the background is dominated by its $t/W$ component (see
Table~\ref{tab:TauCutflow}). In contrast
to the triple-Higgs signal in which the missing energy originates from the two
neutrinos associated with the tau decays, most background events feature either
more than two neutrinos, or a missing energy originating from a $W$-boson pair.
This suggests to take advantage of the $m_{T2}$ variable~\cite{Lester:1999tx,%
Barr:2003rg} to ensure an efficient background rejection. The $m_{T2}$ spectrum
is bounded from above and its shape depends both on a test mass and on the mass
of the semi-invisibly decaying particle. Moreover, the upper bound sharply rises
for increasing test masses above the true mass of the invisible particle~\cite{%
Barr:2007hy}. As the true invisible mass is zero for the triple Higgs signal,
the associated $m_{T2}$ distribution is naturally broader than for the
background,
provided the test mass is taken large enough. This feature is illustrated in
Fig.~\ref{fig:mt2} for which we have chosen an optimized test mass of 190~GeV,
which allows for a maximal background and signal separation.

After having reconstructed all three Higgs bosons, we derive the invariant mass
of the triple-Higgs system $m_{hhh}$ and constrain it to be smaller than
1.6~TeV.

\section{Results and discussion}\label{sec:results}
\begin{table}
  \renewcommand{\arraystretch}{1.1}
  \setlength\tabcolsep{4pt}
  \begin{tabular}{c||c|ccc}
    Selection & Signal & $t/W$ & $X_{\tau\tau}$ & $hh$\\
    \hline \hline
    Baseline & 27 & $1.0 \times 10^{6}$ & $1.6 \times 10^{7}$
      & $1.3 \times 10^{3}$\\
    $m_{\tau \tau}$ & 12 & $1.4 \times 10^{5}$ & $2.6 \times 10^{6}$
      & 670\\
    Boosted Higgs & 0.92 & 640 & $6.5 \times 10^{3}$ & 35 \\
    $m_{\rm j j}$ & 0.47 & 180 & 81                  & 4.1\\
    $b$-tagging   & 0.15 & 15  & 0.20                & 0.034\\
    $m_{T2}$      & 0.11 & 0.37& 0.093               & 0.029\\
    $m_{hhh}$     & 0.10 & $8.5 \times 10^{-3}$ & 0.012 &0.026\\
    \hline \hline
    $S/B$ & \multicolumn{4}{c}{$2.1$}\\
    $\sigma $ & \multicolumn{4}{c}{$2.0$}\\
  \end{tabular}
  \caption{\it Signal and background cross sections, in ab, at different stage
    of the analysis strategy depicted in Sec.~\ref{sec:analysis}. The signal
    to background ratio $S/B$ and the significance $\sigma$ for a luminosity of
    30~ab$^{-1}$ are also indicated.}
  \label{tab:TauCutflow}
\end{table}

We present in Table~\ref{tab:TauCutflow} the fiducial cross sections resulting
from the application of the various selections introduced in
Sec.~\ref{sec:analysis}, both for the signal (assuming the SM case with \mbox{%
$\kappa_3 = \kappa_4 = 0$}) and the background. We can observe the
complementarity of the various steps, the $m_{\tau\tau}$ and boosted Higgs
requirements reducing the background by a factor of more than 2000, while the
reconstruction of the resolved Higgs boson and the $b$-tagging conditions bring
the signal over background ($S/B$) ratio down to the percent level.
The background is at this stage dominated by $t/W$ events and is further reduced
to a manageable level by means of the $m_{T2}$ selection. The selection on
the triple-Higgs invariant mass finally brings the background rate to
half the signal one for the considered benchmark.

In order to set limits and derive the future collider sensitivity in the
$(\kappa_3, \kappa_4)$ plane, we compute a significance $\sigma$ defined as the
likelihood ratio~\cite{Cowan:2010js}
\beq
  \sigma \equiv
    \sqrt{-2\,\ln\bigg(\frac{L(B | S\!+\!B)}{L( S\!+\!B| S\!+\!B)}\bigg)}
  \ \text{with}\
  L(x |n) =  \frac{x^{n}}{n !} e^{-x} \,,
\label{Eq:sigfinicance} \eeq
where $S$ and $B$ are the expected number of signal and background events
respectively. The signal sensitivity turns out to be of about $2\sigma$ in the
SM case for a luminosity of 30~ab$^{-1}$, with a number of signal events
\mbox{$S\sim 3$} and background events \mbox{$B\sim 1.4$}. The number of signal
events could however be increased by considering the strategic approach of
including the contributions of a semi-leptonic
$\tau_h \tau_l b \bar{b} b \bar{b}$ final state, as it
has been recently proposed for di-Higgs searches at the
LHC~\cite{CMS:2015nat}.

\begin{figure*}
  \centering
  \includegraphics[scale=0.60]{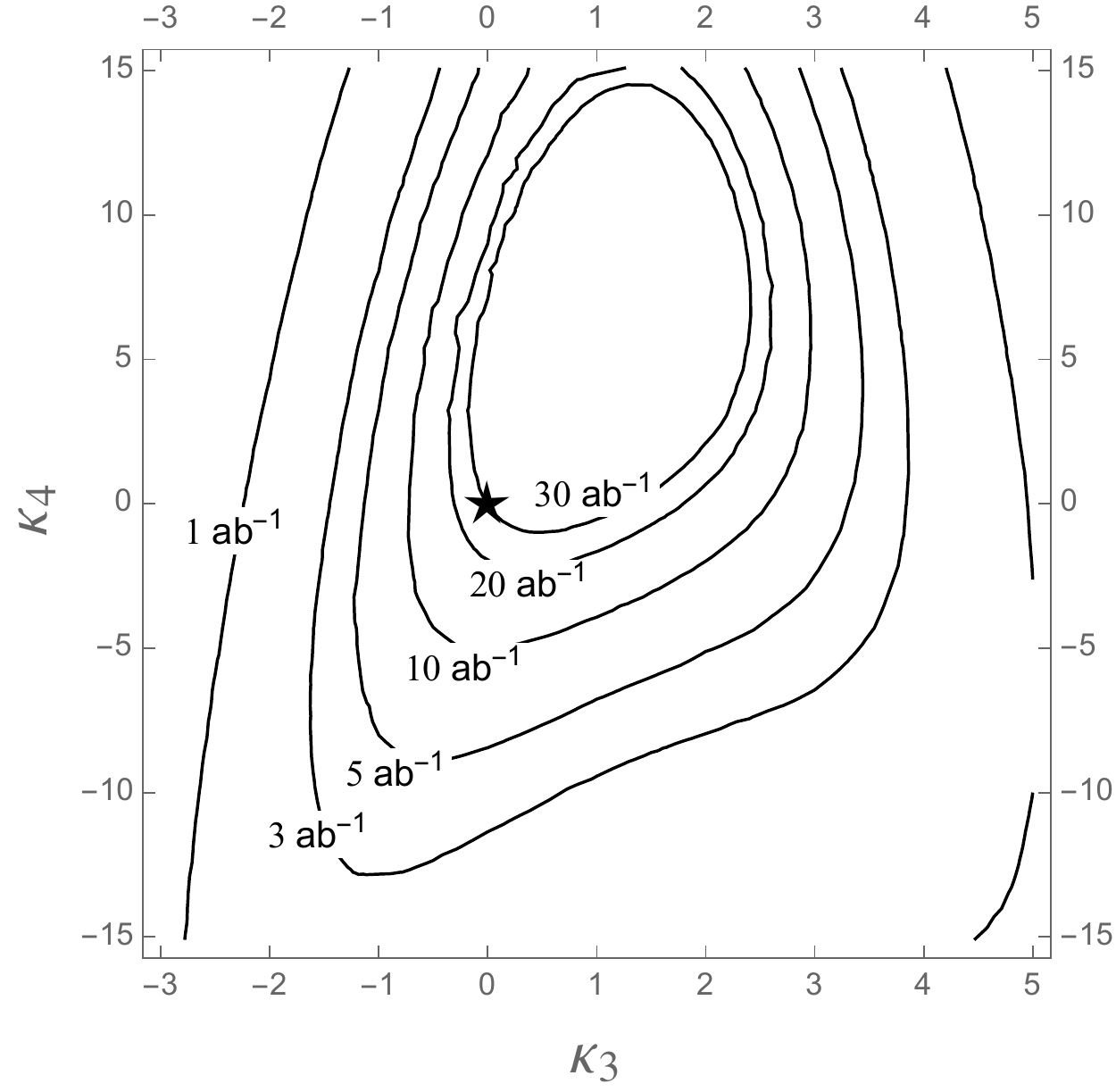}\hspace{0.5cm}
  \includegraphics[scale=0.60]{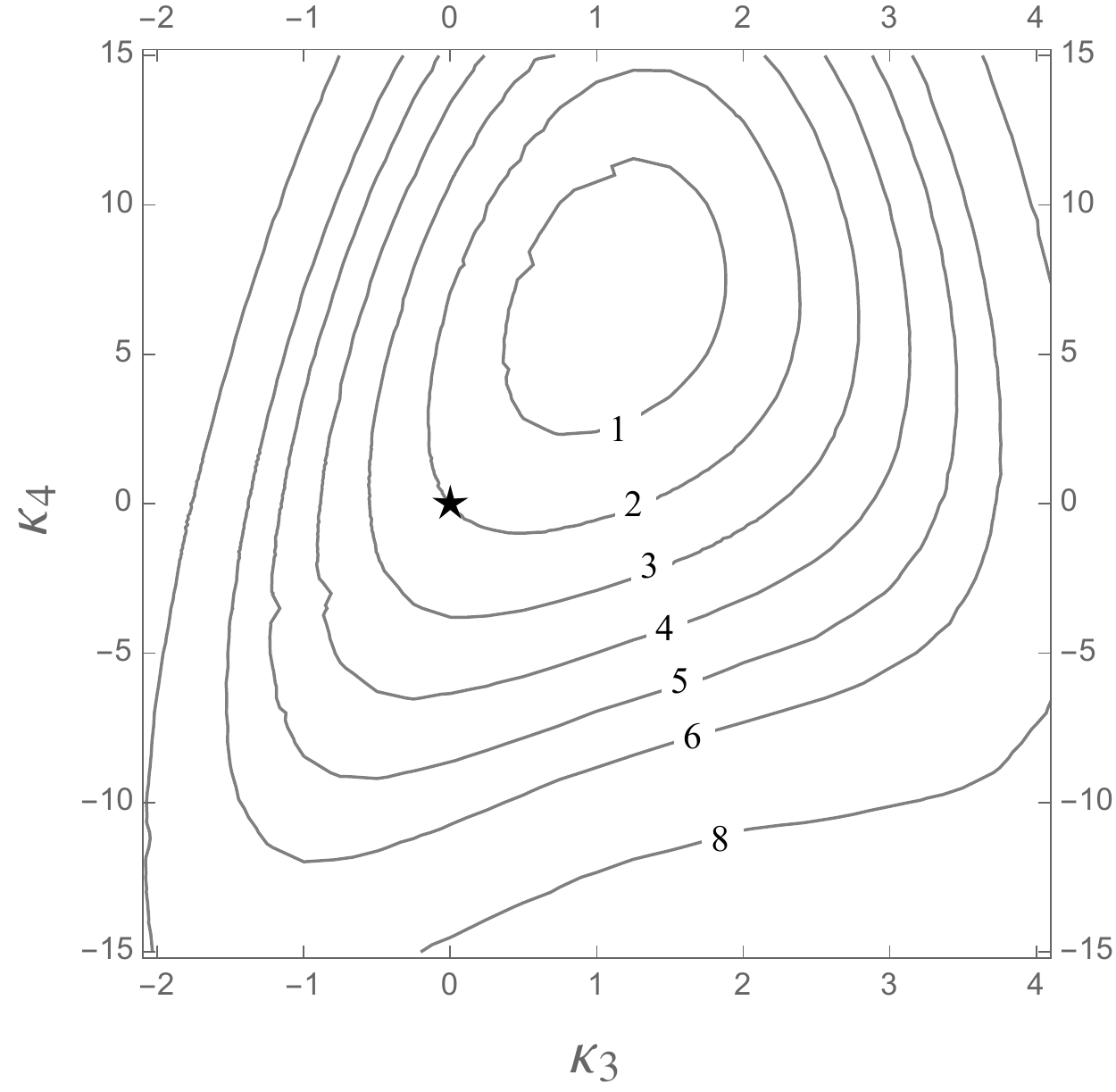}
  \caption{\it Minimum luminosity of 100~TeV proton-proton collisions required
    to achieve a $2\sigma$ sensitivity to a triple-Higgs signal in the
    $b\bar bb\bar b\tau\tau$ channel shown in terms of the $\kappa_3$ and
    $\kappa_4$ parameters (left), and the corresponding sensitivity
    expected for a luminosity of 30~ab$^{-1}$ (right).}
  \label{fig:Exc}
\end{figure*}

Scanning over the $\kappa_i$ parameters, we show in Fig.~\ref{fig:Exc} the
luminosity goals of a 100~TeV proton-proton collider necessary for achieving
a $2\sigma$ exclusion (left panel). Despite the dominance of destructive
interferences on the upper-right-corner of the $(\kappa_3, \kappa_4)$ plane, our
analysis demonstrates that the SM expectation can in principle be excluded with
30~ab$^{-1}$. Conversely, we present in the right panel of the figure the
significance contours obtained when considering a luminosity of 30~ab$^{-1}$.
In order to access the sensitivity gap in the parameter space region limited by
$\kappa_3\in [0, 2]$ and $\kappa_4 \in [0, 14]$, one could combine our results
with other channels, like the $\tau_h \tau_l b \bar{b} b \bar{b}$ mode that
could enhance the sensitivity of the present analysis, and the
$\gamma \gamma b \bar{b} b \bar{b}$ channel investigated in
Refs.~\cite{Fuks:2015hna,Chen:2015gva,Papaefstathiou:2015paa}. Our findings
could moreover be merged with the more precise prospects on the $\kappa_3$
parameters that stem from di-Higgs probes expected to be produced at a large
rate~\cite{Contino:2016spe,Azatov:2015oxa}.

In this work, we have continued our investigation of the possibilities of a
future proton-proton collider expected to run at $\sqrt{s}=100$~TeV
to unravel the true nature of the EWSB mechanism. We have shown that in
addition to the $\gamma \gamma b \bar{b} b \bar{b}$ golden channel, the
$b\bar bb\bar b\tau\tau$ mode is a complementary probe to the quartic Higgs
self-interaction. Our results are comparable to those derived in other
triple-Higgs channels, so that combinations of several searches
could offer handles to parameter space regions featuring
low cross sections and not accessible with a single triple-Higgs analysis.
Such a combination also gives hope to access the
SM couplings beyond the $3\sigma$ level.

\bigskip
\emph{Acknowledgements:} 
We are very grateful to Minho Son for valuable help and discussions, in
particular on tau reconstruction, as well as to K.C.~Kong, Ian M. Lewis and
Graham Wilson for useful comments and suggestions during the course of this
project. We also thank
the HTCaaS group of the Korea Institute of Science and Technology Information
(KISTI) for providing the necessary computing resources and
acknowledge the Korea Future
Collider Study Group (KFCSG) for motivating us to proceed with this work.
JHK is supported in
part by US-DOE (DE-FG02-12ER41809) and by the University of Kansas General
Research Fund allocation 2302091. SL is supported by the National Research
Foundation of Korea (NRF) grant funded by the Korea government (MEST)
(No.~NRF-2015R1A2A1A15052408), and by the Korean Research Foundation
(KRF) through the Korea-CERN collaboration program (NRF-2016R1D1A3B01010529).
The work of BF is partly supported by French state funds managed by the Agence
Nationale de la Recherche (ANR), in the context of the LABEX ILP
(ANR-11-IDEX-0004-02, ANR-10-LABX-63), and by the FKPPL initiative of the CNRS.

\bibliographystyle{JHEP}
\bibliography{draft_paper}

\end{document}